\documentstyle[12pt]{article}
\def\hybrid{\topmargin 0pt      \oddsidemargin 0pt
	\headheight 0pt \headsep 0pt
\textwidth 6.25in       
\textheight 9.5in       
	\marginparwidth .875in
	\parskip 5pt plus 1pt   \jot = 1.5ex}

\catcode`\@=11
\def\marginnote#1{}
\newcount\hour
\newcount\minute
\newtoks\amorpm
\hour=\time\divide\hour by60
\minute=\time{\multiply\hour by60 \global\advance\minute by-\hour}
\edef\standardtime{{\ifnum\hour<12 \global\amorpm={am}%
	\else\global\amorpm={pm}\advance\hour by-12 \fi
	\ifnum\hour=0 \hour=12 \fi
	\number\hour:\ifnum\minute<10 0\fi\number\minute\the\amorpm}}
\edef\militarytime{\number\hour:\ifnum\minute<10 0\fi\number\minute}

\def\draftlabel#1{{\@bsphack\if@filesw {\let\thepage\relax
   \xdef\@gtempa{\write\@auxout{\string
      \newlabel{#1}{{\@currentlabel}{\thepage}}}}}\@gtempa
   \if@nobreak \ifvmode\nobreak\fi\fi\fi\@esphack}
	\gdef\@eqnlabel{#1}}
\def\@eqnlabel{}
\def\@vacuum{}
\def\draftmarginnote#1{\marginpar{\raggedright\scriptsize\tt#1}}

\def\draft{\oddsidemargin -.5truein
	\def\@oddfoot{\sl preliminary draft \hfil
	\rm\thepage\hfil\sl\today\quad\militarytime}
	\let\@evenfoot\@oddfoot \overfullrule 3pt
	\let\label=\draftlabel
	\let\marginnote=\draftmarginnote
   \def\@eqnnum{(\theequation)\rlap{\kern\marginparsep\tt\@eqnlabel}%
\global\let\@eqnlabel\@vacuum}  }

\def\numberbysection{\@addtoreset{equation}{section}
	\def\theequation{\thesection.\arabic{equation}}}

\def\underline#1{\relax\ifmmode\@@underline#1\else
	$\@@underline{\hbox{#1}}$\relax\fi}

\def\titlepage{\@restonecolfalse\if@twocolumn\@restonecoltrue\onecolumn
     \else \newpage \fi \thispagestyle{empty}\c@page\z@
	\def\thefootnote{\fnsymbol{footnote}} }

\def\endtitlepage{\if@restonecol\twocolumn \else  \fi
	\def\thefootnote{\arabic{footnote}}
	\setcounter{footnote}{0}}  
\catcode`@=12
\relax

\def\beq{\begin{equation}}
\def\eeq{\end{equation}}
\def\bea{\begin{eqnarray}}
\def\eea{\end{eqnarray}}

\relax
\hybrid
\begin{document}
\begin{titlepage}
\setcounter{page}{0}
\begin{center}
\hfill hep-lat/9404002

\vskip .2in

{\large CRITICAL POINT CORRELATION FUNCTION FOR THE 2D RANDOM
BOND ISING MODEL}\\[.2in]

        \large  A.L.Talapov\footnote{E-mail addresses:
         talapov@itp.chg.free.net and talapov@pib1.physik.uni-bonn.de}\\

         \normalsize {\it Landau Institute for Theoretical
         Physics\\
         GSP-1 117940 Moscow V-334, Russia}\\

and \\
         \normalsize {\it
         Physikalisches Institut der Universit\"at Bonn\\
         Nussallee 12, 53115 Bonn, Germany}\\[.1in]

        \large   L.N.Shchur\footnote{E-mail address:
         shchur@itp.chg.free.net}\\

         \normalsize {\it Landau Institute for Theoretical
         Physics\\
         GSP-1 117940 Moscow V-334, Russia}\\[.1in]

\vskip .2in

PACS. 05.50 -- Lattice theory and statistics; Ising problems\\
PACS. 75.10 -- General theory and models of magnetic ordering

\vskip .2in

\end{center}

\vskip .2in
\centerline{ ABSTRACT}
\begin{quotation}

High accuracy Monte Carlo simulation results for 1024*1024 Ising system
with ferromagnetic impurity bonds are presented.
Spin-spin correlation function at a critical point
is found to be numerically very close to that of a pure system.
This is not trivial since a critical temperature for the system
with impurities is almost two times lower than pure Ising $T_c$.
Finite corrections to the correlation function due to combined
action of impurities and finite lattice size are described.

\end{quotation}

\end{titlepage}

\newpage

Influence of impurities on a critical behaviour has been a subject of
numerous papers.

For the simplest possible model -- 2D Ising model this problem has been
considered theoretically [1-9],
experimentally [10,11] and using computer simulations
[12-16].

Most of the simulations were devoted to thermodynamic properties such as
specific heat, magnetization and magnetic susceptibility. On the other
hand, theories  give direct predictions for spin-spin correlation
function
\beq
< S(0)S(r) >
\label{L1}
\eeq
where $r$ is a distance between spins.

We used cluster algorithm special purpose processor (SPP) [17,18]
to get accurate
values of $<S(0)S(r)>$ at a critical point.
The SPP realizes in hardware Wolff cluster algorithm, and therefore does not
suffer of critical slowing down.

The SPP spends $375 ns$ per one cluster spin.
It also has a simple hardware for
fast calculation of spin-spin correlation functions. Time necessary to get
the correlation function for some $r$ is equal to $L^2 * 21 ns$,
where $L$ is a linear lattice size.

We study the following model. Coupling constant $J$ on each bond can take two
positive values: $J_1$ with probability $p$ and $J_0$ with probability $1-p$.
For $p=0.5$ duality relation [19] shows that $T_c$ is equal
to that of a pure
model with all horizontal bonds equal to $J_1$ and all vertical bonds equal
to $J_0$. The known $T_c$ greatly simplifies simulation data analysis, and
therefore we used
$p=0.5$.

Theoretical models employ a small parameter
\beq
g \sim p(J_0 - J_1)^2
\label{L2}
\eeq
This parameter is connected with impurity induced length $l_i$
\beq
\log l_i \sim \frac{1}{g}
\label{L3}
\eeq
To be able to notice deviations from the pure critical
behaviour we should have
\beq
l_i << L
\label{L4}
\eeq
To satisfy this condition we used large values of $L$
($L=256,\; 512$ and $1024$) and quite different
values of $J$:
$J_0 =1$ and $J_1 =0.25$. So, in the simulations $g$ is not very small.

Theories deal with continuum limit and infinite lattice size.
Simulations are conducted on a finite lattice with
periodic boundary conditions.
We calculated $<S(0)S(r)>$ for spins, located along one lattice row. In this
case distance $r$ can take only integer values.

Our simulations for the pure case [18] showed that
discrete lattice effects
are significant for $r < 8$. Continuum theory can be applied for larger
distances. But the finite size corrections for $r > 8$ are very significant
and should be taken into account explicitly.

Pure Ising correlation function $c_{0}(r)$ for $r/L \rightarrow 0$ has been
calculated in [20]
\beq
c_{0}(r)=  \frac{0.70338}{r^{1/4}}
\label{L5}
\eeq
Continuum limit of (1) for the finite lattice with periodic boundary
conditions $c(r,L)$ has been obtained in [21]
\beq
c(r,L) \sim \frac{\sum_{\nu =1}^4 \left|\theta _\nu (\frac{r}{2L}) \right|}
{\left|\theta _1 (\frac{r}{L})\right|^{1/4}}
\label{L6}
\eeq
where $\theta$ are Jacobi theta functions.
For our purposes c(r,L) can be written in a simpler form
\beq
c(r,L) \approx A(L) \frac{1+e^{-\pi /4}\left[\sin(\frac{\alpha}{2})+
\cos(\frac{\alpha}{2})\right]
+e^{-9\pi /4}\left[\cos(\frac{3\alpha}{2})-\sin(\frac{3\alpha}{2})\right]}
{\left(\sin(\alpha)-e^{-2\pi}\sin(3\alpha)\right)^{1/4}}
\label{L7}
\eeq
where $\alpha= \pi r/L$.
The coefficient $A(L)$ can be obtained using the expression for $c_{0}(r)$.
Formula for $c(r,L)$ is in excellent agreement with simulation
results [18] for the pure system.

In Fig.1 we show the ratio of computed $<S(0)S(r)>$ to $c(r,L)$ for
$L=256,\; 512$ and $1024$.

For each lattice size
to get mean values of $<S(0)S(r)>$ and standard deviations we used
1000 samples with different impurities distribution.
For each sample all spins initially were pointing in the
same direction. Two thousand Wolff clusters were flipped to thermalize
spin distribution at critical temperature. Another 8000
clusters were flipped to calculate mean values of $<S(0)S(r)>$ for each
sample. For $L=1024$ one cluster flip at $T_c$ requires about $0.1 sec$.

For each lattice size
correlation functions for different $r$ were measured for the same spin
configurations.

Standard errors are determined mainly by different behaviour of
$<S(0)S(r)>$ for
different impurity distributions
and not by thermal fluctuations for a given sample.

Deviations of $<S(0)S(r)>$ from $c(r,L)$ at $r<8$ are due to the discrete
lattice effects. In fact, at these distances correlation function of the
impure system is extremely close to the pure correlation function. For
$r=1$ difference between them is of the order of $10^{-3}$.
At first sight it seems
to be quite natural, because $r<<l_{i}$ and $<S(0)S(0)>=1$. But the critical
temperatures for pure ($1/T_c\approx 0.44068679$)
and impure ($1/T_c\approx 0.80705186$)
cases are very different, and continuum theory
cannot exclude strong renormalization of $<S(0)S(1)>$.

On the other hand, for $r>8$ pure correlation function practically
coincides with $c(r,L)$. So, Fig.1 shows that impurities decrease
spin-spin correlations at large $r$. Again, this is not trivial because
of the difference of critical temperatures for pure and impure cases.
Moreover, we see from Fig.1, that for a given $r$, the larger the lattice
size $L$, the smaller deviations from the pure behaviour.

To investigate this phenomenon more, in Fig.2 we draw the same data
for $L=256,512,1024$ as a function of $(r/L)$. We see, that $L=256$
and $L=512$ data practically coincide. The difference between these data
and $L=1024$ data is also within the limits of the error bars.

We are forced to conclude that the influence of impurities can be described
by some function $F(g,(r/L))$. The impure system correlation function
for any $L$ is just a product of $F(g,(r/L)$ and $c(r,L)$.
The behaviour of $F$ as a function of $(r/L)$ is given by Fig.2.
Even in the case of rather strong impurities, which we simulated using
the SPP, $F$ is very close to $1$ for small $(r/L)$, and decreases
only by 4 percents from $1$ at largest possible value of $(r/L)=0.5$.


These results are in contradiction with predictions of
[22] for the spin--spin correlation function, which would instead show
increase in spin correlations at large distances due to impurities.
First order renormalization group
calculations of other authors [2-6] lead to the conclusion that there
is no influence of
impurities on the averaged spin--spin correlation function.

Our results show that there exist unexpectedly small
finite renormalization of
the critical point correlation function.
Our simulation data cannot exclude third order
renormalization group corrections, recently found to be nonzero [23].
These corrections should somewhat increase the correlation function
at large $r$.

\vspace {0.5truecm}
\noindent{\large Acknowledgements}

This work was started in collaboration with Vl.S.Dotsenko, and we are
especially grateful to him.
We are indebted to R.Balian for helpful suggestion.
We also acknowledge useful discussions with Vik.S.Dotsenko,
V.L.Pokrovsky, W.Selke, A.Compagner,
H.W.J.Blote, J.R.Herringa, A.Hoogland and I.T.J.C.Fonk.

This work is partially supported by grants 07-13-210 of NWO,
the Dutch Organization of Scientific Research and 93-02-2018 of RFFR,
the Russian Foundation for Fundamental Research.

\newpage



\newpage

{\bf Figure Captions}

\vspace {1.5truecm}

Fig.1.
Ratio of $<S(0)S(r)>$ to $c(r,L)$.
Dashed line shows data for the pure $L=1024$ Ising
model [18] at $(1/T_c)=.4406868$.
Solid lines connect data points
for the systems with impurities at $(1/T_c)=.8070519$.
Up triangles show $L=1024$ data, empty squares $L=512$ data,
and circles $L=256$ data.

\vspace {1.5truecm}

Fig.2.
The same data as in Fig.1, but shown as a function of $(r/L)$.

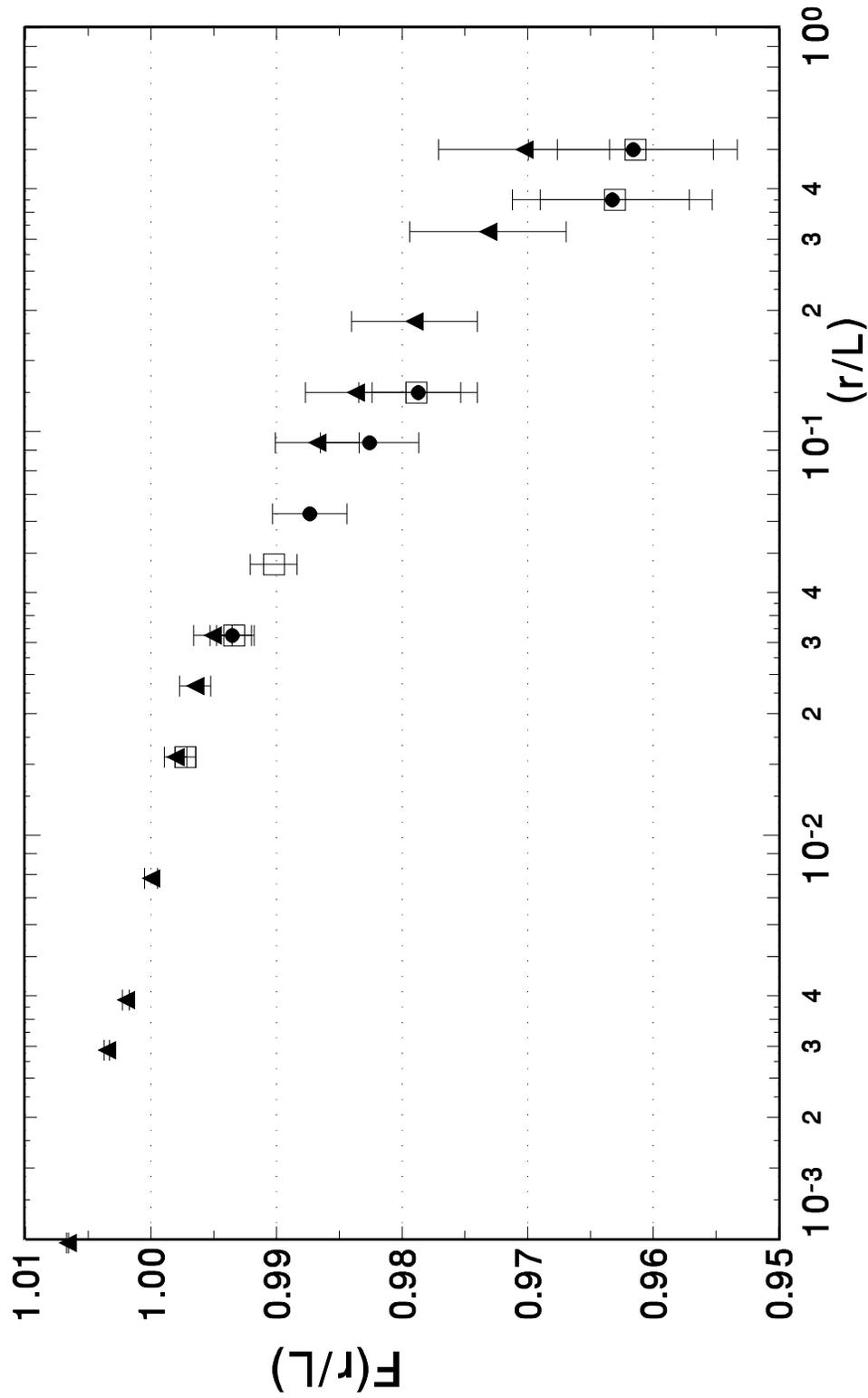
\end{document}